# DFT Investigation of Biocatalytic Mechanisms from pH-Driven, Multi-Enzyme, Biomimetic Behavior in CeO$_2$

Hongyang Ma,*,[a] Zhao Liu,[b] Pramod Koshy,[a] Charles C. Sorrell,[a] and Judy N. Hart*,[a]

[a] School of Materials Science and Engineering, UNSW Sydney, Sydney, NSW 2052, Australia

[b] Sino-French Institute of Nuclear Engineering and Technology, Sun Yat-sen University, Zhuhai 519082, China

*Corresponding Authors: hongyang.ma@unsw.edu.au, j.hart@unsw.edu.au

(1)

**Abstract**

There is considerable interest in the pH-dependent, switchable, biocatalytic properties of cerium oxide ($CeO_2$) nanoparticles (CeNPs) in biomedicine, where these materials exhibit beneficial antioxidant activity against reactive oxygen species (ROS) at basic physiological pH but cytotoxic prooxidant activity in acidic cancer cell pH microenvironment. While the general characteristics of the role of oxygen vacancies are known, the mechanism of their action at the atomic scale under different pH conditions has yet to be elucidated. The present work applies density functional theory (DFT) calculations to interpret, at the atomic scale, the pH-induced behavior of the stable {111} surface of $CeO_2$ containing oxygen vacancies. Analysis of the surface-adsorbed media species reveals the critical role of pH on the interaction between ROS ($^\bullet O_2^-$ and $H_2O_2$) and the defective $CeO_2$ {111} surface. Under basic conditions, the superoxide dismutase (SOD) and catalase (CAT) biomimetic reactions can be performed cyclically, scavenging and decomposing ROS to harmless products, making $CeO_2$ an excellent antioxidant. However, under acidic conditions, the CAT biomimetic reaction is hindered owing to the limited reversibility of $Ce^{3+} \leftrightarrow Ce^{4+}$ and formation $\leftrightarrow$ annihilation of oxygen vacancies. A Fenton biomimetic reaction ($H_2O_2 + Ce^{3+} \rightarrow Ce^{4+} + OH^- + \,^\bullet OH$) is predicted to occur simultaneously with the SOD and CAT biomimetic reactions, resulting in the formation of hydroxyl radicals, making $CeO_2$ a cytotoxic prooxidant.

**Keywords:** $CeO_2$, pH, ROS, Oxygen Vacancies, DFT

**Introduction**

Oxidative damage promoted by reactive oxygen species (ROS) is a primary cause or aggravation of many human pathologies, including cancer[1]. Accordingly, considerable effort has been made to develop antioxidant therapies that can counter the harmful effects driven by ROS, protect cells against oxidative damage and related pathologies, and treat diseases caused by oxidative damage[2-4]. In this context, $CeO_2$ nanoparticles (CeNPs) have attracted considerable attention as a biocompatible antioxidant tool owing to their ability to eliminate ROS efficiently, thus protecting cell activities and allowing their long-lasting performance in regulating redox metabolism[5-8].

The biomedical activity of CeNPs is attributed primarily to the rapidly reversible switching of $Ce^{3+} \leftrightarrow Ce^{4+}$ and the respective formation $\leftrightarrow$ annihilation of oxygen vacancies ($V_O^{\bullet\bullet}$), which confer on CeNPs a unique auto-regenerative antioxidant property[9-13]. This property has placed CeNPs in the spotlight owing to the efficient ROS-scavenging effect through a combination of: (1) a superoxide dismutase (SOD) biomimetic reaction, where the superoxide ions ($^{\bullet}O_2^-$) are reduced to peroxide groups by $Ce^{3+} \rightarrow Ce^{4+}$ oxidation[14], and (2) a catalase (CAT) biomimetic reaction, where hydrogen peroxides ($H_2O_2$) are decomposed to $H_2O$ and $O_2$ by the reverse $Ce^{4+} \rightarrow Ce^{3+}$ reduction[15]:

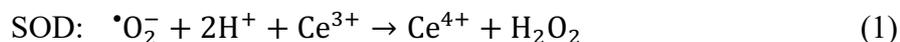
$$\text{SOD:} \quad {}^{\bullet}O_2^- + 2H^+ + Ce^{3+} \rightarrow Ce^{4+} + H_2O_2 \qquad (1)$$

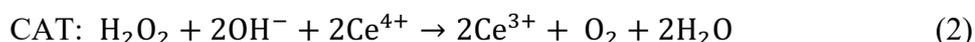
$$\text{CAT:} \quad H_2O_2 + 2OH^- + 2Ce^{4+} \rightarrow 2Ce^{3+} + O_2 + 2H_2O \qquad (2)$$

Thereby, CeNPs can undergo a reversible biocatalytic redox cycle, eliminating the toxic ROS (*i.e.*, $^{\bullet}O_2^-$ and $H_2O_2$) while being returned to their original state[13]. As a result, CeNPs have received widespread attention for biomedical applications including drug delivery[16-18], bioanalysis[19-21], and bioscaffolding[22]. An increasing range of multifunctional nanoparticles has been developed through the surface modification of CeNPs and the integration of CeNPs into other biomaterials in order to enhance the regulation of ROS and their therapeutic effects[23-26].

Of particular interest is the pH-dependent biocatalytic performance of CeNPs, which results in beneficial antioxidant activity against ROS and promotion of cell proliferation at basic physiological pH, while they facilitate cytotoxic prooxidant activity and induce cell apoptosis at acidic pH[18, 27, 28]. This unique behavior makes CeNPs a potential tool to treat cancer since cancer cells have a more acidic cytosolic pH than normal cells due to the Warburg effect[29]. However, the use of CeNPs in biomedicine remains controversial since the mechanisms associated with the selective antioxidant and prooxidant behavior under different pH conditions are ill-defined and poorly understood[10]. Additionally, there remains some exceptions that bring into question the universality of the selective cytotoxicity of CeNPs, such as the observation of normal cell apoptosis[30] and cancer cell protection[13, 31, 32]. This ambiguity currently limits the potential of CeNPs to laboratory and research settings and prevents clinical applications. Therefore, further efforts are required in order to gain a greater understanding of the mechanisms associated with the biocatalytic properties of CeNPs for cancer therapies.

The differential sensitivity of the SOD and CAT biomimetic reactions of CeNPs to pH conditions has been suggested as one of the primary factors resulting in the selective toxicity of CeNPs[5, 10, 33]. Although the interaction of oxygen molecules and ROS with pristine low-index $CeO_2$ surfaces has previously been investigated[34-38], there do not appear to be any comprehensive studies that consider the effects of pH and $V_O^{\bullet\bullet}$ on these processes. Prior work by the authors has shown that media molecules and pH conditions have a significant influence on the reversibility of the state of $V_O^{\bullet\bullet}$[39]. The present work uses density functional theory (DFT) calculations to reveal, at the atomic scale, the influence of pH on the interaction of $^{\bullet}O_2^-$ (*i.e.*, the SOD biomimetic reaction) and $H_2O_2$ (*i.e.*, the CAT biomimetic reaction) with the most stable $CeO_2$ {111} surface. These data may provide the first step in better control of the pH-induced behavior of CeNPs, leading to their use for the treatment of cancer and other diseases.

**Computational Methods**

Spin-polarized periodic density functional theory (DFT) calculations were employed to reveal the influence of pH on the interaction of $^{\bullet}O_2^-$ and $H_2O_2$ with the defective $CeO_2$ {111} surface. The computational methods are built on those used in the authors' previous work[39]. Briefly, the calculations were performed with the CRYSTAL17 code[40] using a B3LYP-like method[41, 42] to calculate the exchange-correlation energy. 12% of Hartree-Fock (HF) exchange energy mixed with the DFT exchange-correlation was used. The van der Waals interactions were included in all calculations, with C6 coefficients chosen according to the Grimme D3 scheme[43]. An effective core pseudopotential for cerium[44], an 8-411(d1) basis set for oxygen[45], and a TZVP basis set for hydrogen[46] were used. The Coulomb and exchange series were truncated with threshold values of $10^{-7}$, $10^{-7}$, $10^{-7}$, $10^{-7}$, and $10^{-14}$. The convergence threshold for energy for the self-consistent-field (SCF) procedure was set to $1 \times 10^{-7}$ Hartree (Ha). For the geometry optimization, the convergence criteria for the root mean square (RMS) force and RMS atomic displacement values were set to $3 \times 10^{-4}$ Ha/Bohr and $1.2 \times 10^{-3}$ Bohr, respectively, while the convergence criteria for the maximal force and maximal atomic displacement were 1.5 times their respective RMS values.

A defective {111} surface model was constructed identical to that of the previous work[39], with a thickness of three O−Ce−O trilayers, with the atoms in the bottom O−Ce−O trilayer frozen in their bulk positions. A 3 × 3 slab supercell, providing a ∼ 12 × 12 Å$^2$ surface area, was used to avoid interactions between periodic images of the $V_O^{\bullet\bullet}$ and ROS. The *k*-point meshes were sampled using a 2 × 2 × 1 Monkhorst-Pack grid for the supercell model of the CeO$_2$ {111} slab structure; convergence with respect to the number of *k*-points was checked (Figure S1, Supporting Information). The Ce$^{3+}$ that result from the creation of a $V_O^{\bullet\bullet}$ in the CeO$_2$ surface localize on the NN-NNN sites (NN: nearest neighbor Ce to the $V_O^{\bullet\bullet}$; NNN: next-nearest neighbor Ce to the $V_O^{\bullet\bullet}$). Ten aqueous-phase media molecules, including nine H$_2$O and one H$_3$O$^+$/OH$^−$ ion for acidic/basic conditions, were placed above the CeO$_2$ {111} surface to simulate different pH conditions, and the positions of the molecules and surface atoms were optimized. A uniform background charge was added in order to neutralize the charges (*i.e.*, of the H$_3$O$^+$ or OH$^−$ ion) in the simulation cell. For acidic conditions, a proton (H$^+$) is released from the H$_3$O$^+$ ion during optimization and adsorbs on the surface oxygen. Additionally, based on the results from the previous work[39], the defective {111} surface of CeO$_2$ behaves differently under different pH conditions. Under acidic conditions, the $V_O^{\bullet\bullet}$ is filled partially by an H$_2$O molecule, while, under neutral and basic conditions, the $V_O^{\bullet\bullet}$ is filled and annihilated completely by the OH$^−$ ion. After this optimization, one $^\bullet$O$_2^−$ or H$_2$O$_2$ was introduced either near to or far from the $V_O^{\bullet\bullet}$ in order to elucidate the influence of pH on the SOD and CAT biomimetic reactions. In order to interpret the contribution of the $V_O^{\bullet\bullet}$ to the interaction between the ROS ($^\bullet$O$_2^−$ and H$_2$O$_2$) and the CeO$_2$ {111} surface, two different cases were considered at each pH: (1) the $V_O^{\bullet\bullet}$ site was filled initially by an H$_2$O molecule (acidic condition) or an OH$^−$ ion (basic condition), labeled as VFG (vacancy-filled geometry), and (2) the $V_O^{\bullet\bullet}$ was vacant initially, with the ions determining the pH (H$_3$O$^+$ or OH$^−$) being placed near the $V_O^{\bullet\bullet}$, labeled as VUG (vacancy-unfilled geometry) (Figure S2, Supporting Information).

Multiple initial geometries with different arrangements of the media molecules and ROS were tested for each pH condition and initial vacancy state (*i.e.*, VFG and VUG). The adsorption energies of the ROS on the defective CeO$_2$ {111} surface, $\Delta E_{ads}$, were calculated by:

$$\Delta E_{ads} = E_{slab/media/ROS} - E_{slab/media} - E_{ROS} \tag{3}$$

where $E_{slab/media/ROS}$ is the energy of the defective slab with adsorbed media molecules and ROS, $E_{slab/media}$ is the energy of the defective slab with adsorbed media molecules but without ROS, and $E_{ROS}$ is the energy of the isolated ROS.

**Results and Discussion**

**Influence of pH on SOD Biomimetic Reaction**

To elucidate the influence of different pH conditions on the activity of the SOD biomimetic reaction, various spatial arrangements of aqueous-phase media molecules and ROS above the defective $CeO_2$ {111} surface containing two $Ce^{3+}$ were tested (Figures S3 and S4, Supporting Information). It is notable that, in all of the optimized geometries of both VFG and VUG, the initial $^•O_2^-$ is reduced spontaneously to a peroxide group ($O_2^{2-}$) by electron transfer from $Ce^{3+}$ to $^•O_2^-$, resulting in one $Ce^{3+}$ switching to $Ce^{4+}$; the resultant peroxide is adsorbed on a surface Ce. The adsorption energies shown in Table 1 are for the combined spontaneous process of $^•O_2^-$ adsorption and electron transfer from $Ce^{3+}$ to $^•O_2^-$.

**Table 1.** Energies for adsorption of $^•O_2^-$ and spontaneous electron transfer from $Ce^{3+}$ to $^•O_2^-$ on the defective $CeO_2$ {111} surface under different pH conditions (VFG = vacancy-filled geometries; VUG = vacancy-unfilled geometries)

|  | Adsorption Energy (kcal/mol) | | | |
|---|---|---|---|---|
|  | Acidic | | Basic | |
|  | VFG | VUG | VFG | VUG |
| Geometry 1 | -78.2 | -102.6 | -310.1 | -312.0 |
| Geometry 2 | -59.5 | - 96.4 | -301.0 | -307.4 |

*Basic:* The negative adsorption energies for both the VFG and VUG cases under basic conditions indicate that the adsorption of $^•O_2^-$ and electron transfer are energetically favorable (Table 1). There is negligible variation in the energies for the four different geometries of the media molecules and ROS considered, demonstrating that the spatial arrangement of the media molecules and ROS has an insignificant effect on the interaction between the $^•O_2^-$ and the $CeO_2$ surface.

In the VFG cases, the $V_O^{••}$ remains filled by the $OH^-$ ion after optimization, while, for the VUG cases, the $^•O_2^-$ completely fills the $V_O^{••}$ and the $OH^-$ ion is adsorbed on a surface Ce (Figure 1). Additionally, in the VUG cases, after the accommodation of $^•O_2^-$ in the $V_O^{••}$ and the shift of one Ce from valence 3+ to 4+, the residual surface $Ce^{3+}$ spontaneously migrates from the surface layer (1st-layer) to the subsurface layer (2nd-layer) (Figure 1). This indicates that a lone $Ce^{3+}$ is more favorably located in the 2nd-layer rather than 1st-layer. However, in the VFG cases, the spontaneous migration of the residual surface $Ce^{3+}$ is not observed, with the $Ce^{3+}$ being located adjacent to the protonated surface oxygen (*i.e.*, the $OH^-$ ion that initially is accommodated in the $V_O^{\bullet\bullet}$). This presumably is due to the existence of the proton on the $CeO_2$ surface, which stabilizes the surface $Ce^{3+}$ (Figure 1).

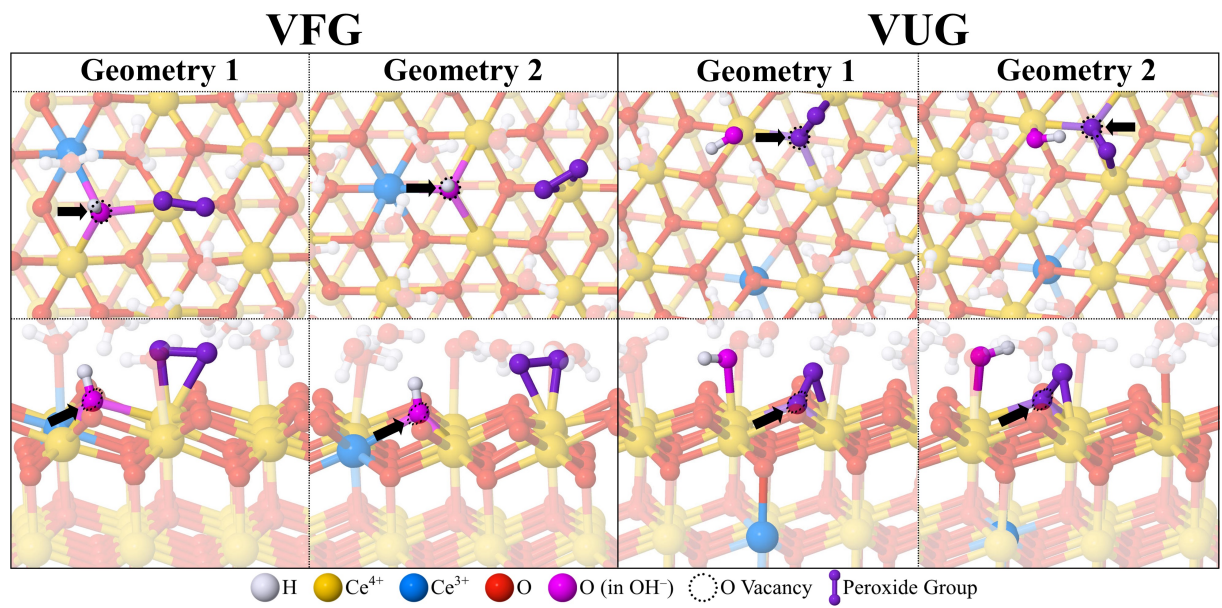

**Figure 1.** Optimized vacancy-filled geometries (VFG) and vacancy-unfilled geometries (VUG) with different arrangements of media molecules and ROS ($^\bullet O_2^-$) adsorbed on the defective $CeO_2$ {111} surface (*i.e.*, containing an oxygen vacancy ($V_O^{\bullet\bullet}$)) initially with 2 $Ce^{3+}$ under basic conditions (top views in upper row; corresponding side views in lower row); black arrows indicate locations of $V_O^{\bullet\bullet}$

*Acidic:* Compared to the basic pH cases, significantly less negative adsorption energies indicate that the adsorption of $^\bullet O_2^-$ and the electron transfer from $Ce^{3+}$ to $^\bullet O_2^-$ are less energetically favorable under acidic conditions (Table 1). Additionally, as opposed to the basic pH cases, the adsorption energies of $^\bullet O_2^-$ are influenced significantly by the initial state of filling of the $V_O^{\bullet\bullet}$ under acidic conditions. In the VUG cases, the $^\bullet O_2^-$ is accommodated completely in the $V_O^{\bullet\bullet}$ after optimization (Figure 2). In the VFG cases, the $V_O^{\bullet\bullet}$ remains filled partially by an $H_2O$ molecule after optimization. The less negative adsorption energies of the VFG cases compared to the VUG cases indicate that the accommodation of $H_2O$ molecules in the $V_O^{\bullet\bullet}$ provides a less stable surface structure than the accommodation of $^\bullet O_2^-$. This can be explained by the complete filling of the $V_O^{\bullet\bullet}$ by $^\bullet O_2^-$ and the formation of three to four strong $O-Ce_{surface}$ bonds with lengths in the range 2.3-2.5 Å. However, the $H_2O$ molecules can only partially fill

(7)

the $V_O^{\bullet\bullet}$, forming a maximum of two O$_{water}$–Ce$_{surface}$ bonds with lengths in the range 2.7-2.9 Å, due to the large molecular size and high initial coordination of the oxygen in H$_2$O.

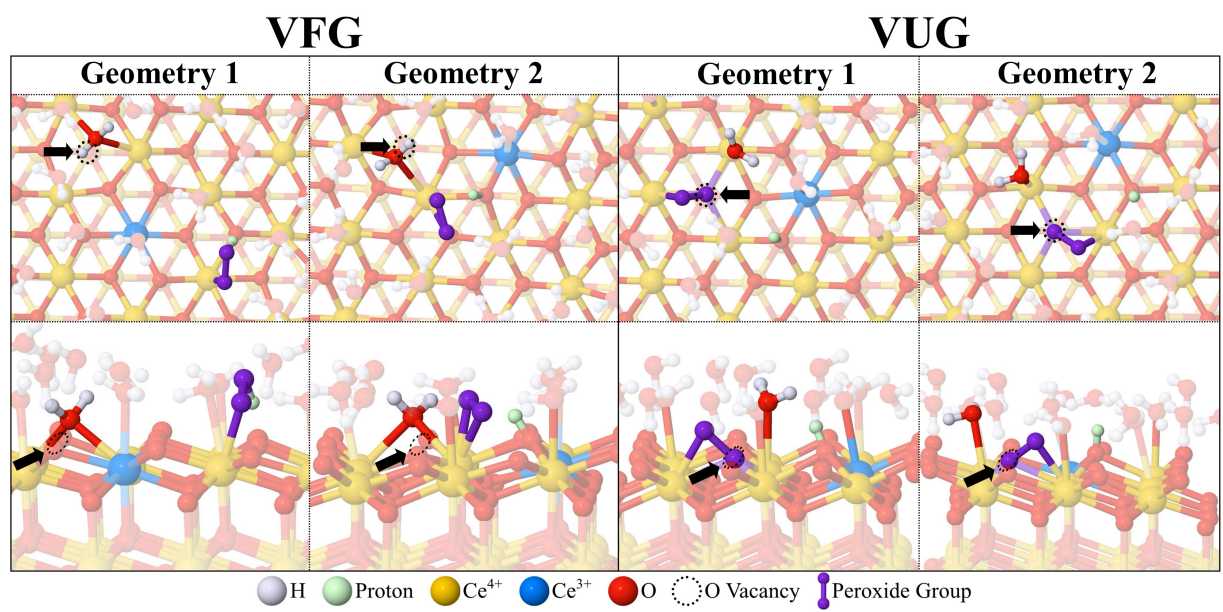

**Figure 2.** Optimized vacancy-filled geometries (VFG) and vacancy-unfilled geometries (VUG) with different arrangements of media molecules and ROS ($^{\bullet}O_2^-$) adsorbed on the defective CeO$_2$ {111} surface (*i.e.*, containing an oxygen vacancy ($V_O^{\bullet\bullet}$)) initially with 2 Ce$^{3+}$ under acidic conditions (top views in upper row; corresponding side views in lower row); black arrows indicate locations of $V_O^{\bullet\bullet}$

Additionally, in Geometry 1 of VFG, an HO$_2^-$ ion spontaneously forms through the interaction between $^{\bullet}O_2^-$ and the proton that was adsorbed initially on the surface (Figure 2). More specifically, when the electron transfer from the surface Ce$^{3+}$ to the $^{\bullet}O_2^-$ occurs, the resultant peroxide group captures the surface proton. However, the occurrence of this process is not seen in Geometry 2 of VFG, indicating that the spontaneous formation of an HO$_2^-$ ion is dependent on the spatial arrangement of the media molecules. The more negative adsorption energy for Geometry 1 indicates that the formation of the HO$_2^-$ ion is more favorable.

In both the VFG and VUG cases under acidic conditions, the spontaneous migration of Ce$^{3+}$ from the 1$^{st}$-layer to the 2$^{nd}$-layer is not observed (Figure 2). This is attributed to the same cause as discussed for the basic pH cases, where the surface-adsorbed proton stabilizes the Ce$^{3+}$ in the 1$^{st}$-layer. In Geometry 1 of VFG (Figure 2), in which no surface proton exists since the initially adsorbed proton is captured by the peroxide group to form HO$_2^-$, the $V_O^{\bullet\bullet}$ is filled only partially by H$_2$O. Since the $V_O^{\bullet\bullet}$ is not annihilated and is retained, this may have a similar effect to that of the protons in stabilizing the Ce$^{3+}$ in the 1$^{st}$-layer.

(8)

***Summary of SOD biomimetic reaction:*** The preceding results indicate that the pH significantly influences the surface adsorption of $^\bullet O_2^-$, which initiates the SOD biomimetic reaction. The rate of completion of the reaction through the subsequent steps is concluded to be influenced primarily by three factors: (1) the adsorption of $^\bullet O_2^-$ on the defective $CeO_2$ {111} surface, (2) the availability of $Ce^{3+}$, and (3) the availability of $H^+$ ions. For the first factor, the significantly more negative adsorption energies for basic pH than acidic pH indicate that the adsorption of $^\bullet O_2^-$ on the defective $CeO_2$ {111} surface is more favorable under basic conditions (Table 1). For the second factor, in terms of the availability of $Ce^{3+}$, the migration of $Ce^{3+}$ from the 1$^{st}$-layer to the 2$^{nd}$-layer under basic conditions can result in the lower availability of surface $Ce^{3+}$ for the SOD biomimetic reaction than under acidic conditions. However, this migration only occurs in the VUG cases. For the third factor, the availability of $H^+$ ions for the SOD biomimetic reaction is derived primarily from two sources: (1) the aqueous environment and (2) the release of protons adsorbed on the surface oxygen for acidic conditions and the release of protons from the $OH^-$ ions that are accommodated in the $V_O^{\bullet\bullet}$ for basic conditions. In terms of the availability of $H^+$ ions from both sources, it is clear that acidic conditions are more advantageous. Therefore, although the adsorption of $^\bullet O_2^-$ under basic conditions is more favorable than under acidic conditions, the other factors (*viz.*, the availability of $Ce^{3+}$ and $H^+$ ions) are likely to limit the SOD biomimetic reaction at basic pH. Overall, the SOD biomimetic reaction rate may remain similar under different pH conditions, which is consistent with previous report[5].

**Influence of pH on CAT Biomimetic Reaction**

Calculations to investigate the CAT biomimetic reaction were done differently from those for the SOD biomimetic reaction. In the SOD biomimetic reaction calculations, there were two $Ce^{3+}$ in the initial structure to simulate the defective $CeO_2$ {111} surface (Figure S5(a), Supporting Information). In the CAT biomimetic reaction calculations, the two $Ce^{3+}$ on the surface of the initial geometry were changed to two $Ce^{4+}$ by removing two electrons (Figure S5(c), Supporting Information) because the SOD biomimetic reactions switch two $Ce^{3+}$ to two $Ce^{4+}$ to reduce two $^\bullet O_2^-$ to two peroxide groups. Thus, the CAT biomimetic reaction has been investigated on the assumption that the SOD biomimetic reaction occurs twice to convert both $Ce^{3+}$ to $Ce^{4+}$.

Similar to the approach used to investigate the SOD biomimetic reaction, the VFG and VUG cases again were investigated to elucidate the influence of the initial state of filling of the $V_O^{\bullet\bullet}$ on the interaction between $H_2O_2$ and the $CeO_2$ {111} surface (Figures S6-S8, Supporting

Information). The energies for the adsorption of $H_2O_2$ on the $CeO_2$ {111} surface under different pH conditions are shown in Table 2. There are only small energy differences between the adsorption energies for acidic and basic conditions, indicating that the energetic favorability of $H_2O_2$ adsorption on the defective $CeO_2$ {111} surface is influenced only slightly by the pH. However, as discussed below, the nature of the interactions between $H_2O_2$ and the surface is considerably different.

**Table 2.** Energies for adsorption of $H_2O_2$ on the defective $CeO_2$ {111} surface under acidic conditions (with 1 $Ce^{3+}$) and basic conditions (with no $Ce^{3+}$) (VFG = vacancy-filled geometries; VUG = vacancy-unfilled geometries)

|  | Adsorption Energy (kcal/mol) | | | |
|  | Acidic | | Basic | |
|  | VFG | VUG | VFG | VUG |
| --- | --- | --- | --- | --- |
| Geometry 1 | -31.6 | -27.1 | -34.7 | -33.3 |
| Geometry 2 | -25.6 | -15.1 | -34.1 | -34.0 |
| Geometry 3 | -- | -60.0 | -- | -- |
| Geometry 4 | -- | -49.4 | -- | -- |

***Basic:*** In the VFG cases, where the $V_O^{\bullet\bullet}$ initially is filled completely by the $OH^-$ ion, in both geometries the $H_2O_2$ is adsorbed directly on the surface Ce (Figure 3), forming one O–$Ce_{surface}$ bond, while the $OH^-$ ion remains in the $V_O^{\bullet\bullet}$.

In contrast, in one case of VUG (Geometry 1, VUG, Figure 3), where the $H_2O_2$ initially is placed directly above the $V_O^{\bullet\bullet}$ and the $OH^-$ ion is placed near the $V_O^{\bullet\bullet}$, the $OH^-$ ion captures one proton from the $H_2O_2$, forming an $H_2O$ molecule. The residual part of the $H_2O_2$, *i.e.*, an $HO_2^-$ ion, is accommodated in the $V_O^{\bullet\bullet}$. In the other case of VUG (Geometry 2, VUG, Figure 3), the $OH^-$ ion is accommodated completely in the $V_O^{\bullet\bullet}$ and the $H_2O_2$ is adsorbed on the surface Ce adjacent to the $V_O^{\bullet\bullet}$, even though the $H_2O_2$ initially was placed closer to the $V_O^{\bullet\bullet}$ than the $OH^-$ ion. This indicates that the accommodation of an $OH^-$ ion in the $V_O^{\bullet\bullet}$ occurs more readily than the accommodation of $H_2O_2$, which is likely to be due to the smaller size and one-fold coordinated oxygen in the $OH^-$ ion, which confers the ability to fill completely the $V_O^{\bullet\bullet}$ and recover the surface structure to a non-defective-like structure.

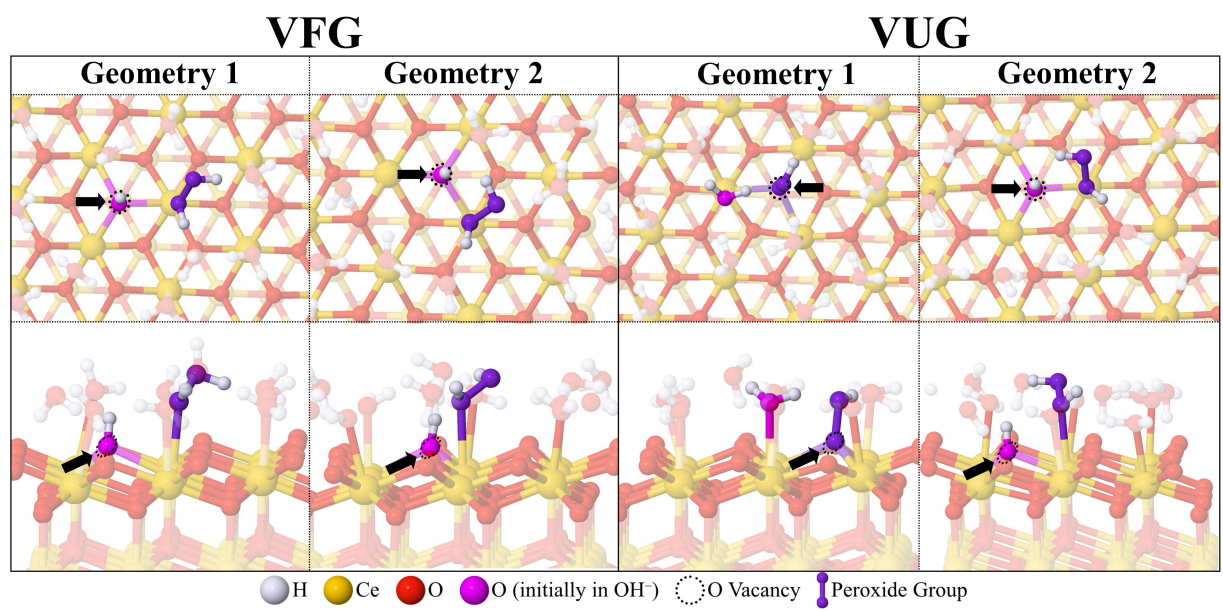

**Figure 3.** Optimized vacancy-filled geometries (VFG) and vacancy-unfilled geometries (VUG) with different arrangements of media molecules and ROS ($H_2O_2$) adsorbed on the defective $CeO_2$ {111} surface (*i.e.*, containing an oxygen vacancy ($V_O^{\bullet\bullet}$)) with no $Ce^{3+}$ under basic conditions (top views in upper row; corresponding side views in lower row); black arrows indicate locations of $V_O^{\bullet\bullet}$

Since the complete CAT biomimetic reaction requires two $OH^-$ ions, a further calculation was done with two $OH^-$ ions placed near the $H_2O_2$ above the defective $CeO_2$ {111} surface in the initial geometry to determine the influence of a higher concentration of $OH^-$ ions on the decomposition of $H_2O_2$ (Figure S9, Supporting Information). After optimization, both O–H bonds in the $H_2O_2$ are broken by the interaction with the two $OH^-$ ions, forming a peroxide group, which is adsorbed in the original $V_O^{\bullet\bullet}$, and two $H_2O$ molecules. Thus, it can be seen that a higher concentration of $OH^-$ ions can facilitate the dissociation of $H_2O_2$ by breakage of the O–H bonds. The higher the concentration of $OH^-$ ions is, the more readily the decomposition of $H_2O_2$ would occur. This decomposition of $H_2O_2$ can facilitate the CAT biomimetic reaction, where the $H_2O_2$ is decomposed to an $O_2$ molecule and the two $H^+$ ions that react with the two $OH^-$ ions to form two $H_2O$. In order to complete the CAT biomimetic reaction, restore the $CeO_2$ surface to its initial state (*i.e.*, a vacant site with two $Ce^{3+}$), and hence be able to repeat the SOD/CAT cycle, the adsorbed peroxide group must be oxidized and released as an $O_2$ molecule. The energy required to release an $O_2$ molecule is calculated to be 4.40 eV, which is slightly less than that to generate a $V_O^{\bullet\bullet}$ by releasing a surface O atom from a pristine $CeO_2$ {111} surface (4.46 eV) (Table S1, Supporting Information). Hence, in terms of the formation of $V_O^{\bullet\bullet}$, the release of an $O_2$ molecule is more favorable than the release of a surface O atom.

(11)

*Acidic:* Surprisingly, under acidic conditions for all four geometries (two geometries for each of the VUG and VFG cases), rather than being adsorbed on the $CeO_2$ surface, some of the media molecules (either $H_2O$ or $H_2O_2$) progressively move away from the $CeO_2$ surface into the vacuum region during the optimization, even though these molecules initially were close to the surface in the initial geometries (Figure S10, Supporting Information). Complete convergence during the optimization procedure could not be achieved.

To understand the origins of this phenomenon, a defective $CeO_2$ surface with no $Ce^{3+}$ and without the introduction of $H_2O_2$ was studied (Figure S11, Supporting Information). A similar phenomenon is seen, where some of the media molecules desorb and escape from the surface. Therefore, the combined results from these five geometries under acidic conditions indicate that the adsorption of media molecules on the defective $CeO_2$ surface with no $Ce^{3+}$ is extremely unfavorable. The origin of these unexpected phenomena observed for these geometries may be that the defective $CeO_2$ {111} surface (*i.e.*, containing a $V_O^{\bullet\bullet}$) with no $Ce^{3+}$ is extremely unstable. Both the $H_2O_2$ and $H_2O$ molecules have similar and relatively large sizes, thereby they can only partially fill the $V_O^{\bullet\bullet}$, so the $V_O^{\bullet\bullet}$ cannot be annihilated to stabilize the surface structure under acidic conditions. This leaves a surface consisting of a $V_O^{\bullet\bullet}$ surrounded by $Ce^{4+}$, which is unstable. This is in contrast to the basic pH cases, where the $V_O^{\bullet\bullet}$ is filled completely and annihilated by the abundant $OH^-$ ions and thus the surface is recovered to a more stable non-defective-like structure.

To find conditions that can promote surface adsorption under acidic conditions, a defective $CeO_2$ surface with one $Ce^{3+}$ was investigated by manually removing only one electron instead of two electrons from the surface (*i.e.*, it is assumed that the SOD biomimetic reaction occurs only once). As expected, in this case, the media molecules are adsorbed on the surface rather than moving away from the surface in the optimized structure (Figure S11, Supporting Information).

The $H_2O_2$ therefore was introduced on a defective surface with only one $Ce^{3+}$. In the VFG cases, where the $V_O^{\bullet\bullet}$ initially is filled partially by an $H_2O$ molecule, the $H_2O_2$ is found above the surface in the optimized structures and forms only weak hydrogen bonds with the surface (Figure S7, Supporting Information). In the VUG cases, the initial arrangement of the media molecules and ROS shows a significant influence on the adsorption of $H_2O_2$ (Figure 4); therefore, to fully investigate this influence, four initial geometries were studied. In Geometries 1 and 2, the $H_2O_2$ is accommodated partially in the $V_O^{\bullet\bullet}$, forming either one or two O–$Ce_{surface}$ bonds with the Ce adjacent to the $V_O^{\bullet\bullet}$. In contrast, in Geometries 3 and 4,

(12)

spontaneous dissociation of the H$_2$O$_2$ occurs by breakage of the O–O bond, resulting in the formation of two hydroxyl radicals (•OH). One •OH captures an electron from the surface Ce$^{3+}$, forming an OH$^-$ ion and shifting the Ce$^{3+}$ to Ce$^{4+}$ (Figure 5). This OH$^-$ ion is accommodated in the $V_O^{\bullet\bullet}$, thus recovering the surface structure to a non-defective-like surface structure, as under basic conditions. The other •OH is retained since no extra Ce$^{3+}$ is available to reduce it to an OH$^-$ ion; the residual •OH is either adsorbed on a surface Ce adjacent to the $V_O^{\bullet\bullet}$ (Geometry 3, Figure 4) or located above a surface Ce (Geometry 4, Figure 4). The more negative adsorption energies for Geometries 3 and 4 (Table 2) indicate that the dissociation of H$_2$O$_2$ is more favorable than molecular (*i.e.*, associative) adsorption. This can be attributed to the resultant formation of an OH$^-$ ion and the subsequent annihilation of the $V_O^{\bullet\bullet}$, which can significantly enhance the stability of the surface structure. The optimized structures formed when H$_2$O$_2$ dissociates under acidic conditions are similar to those found under basic conditions, although the by-product •OH and an additional proton adsorbed on the surface also are present under acidic conditions.

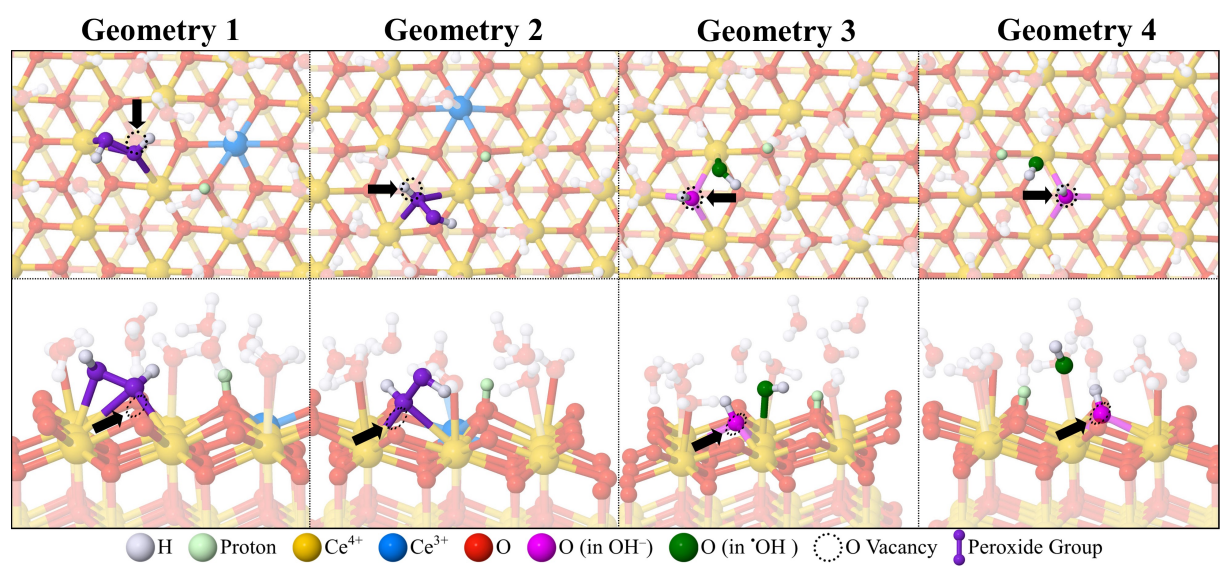

**Figure 4.** Optimized vacancy-unfilled geometries (VUG) with different arrangements of media molecules and ROS (H$_2$O$_2$) adsorbed on the defective CeO$_2$ {111} surface (*i.e.*, containing an oxygen vacancy ($V_O^{\bullet\bullet}$)) initially with 1 Ce$^{3+}$ under acidic conditions (top views in upper row; corresponding side views in lower row); black arrows indicate locations of $V_O^{\bullet\bullet}$

(13)

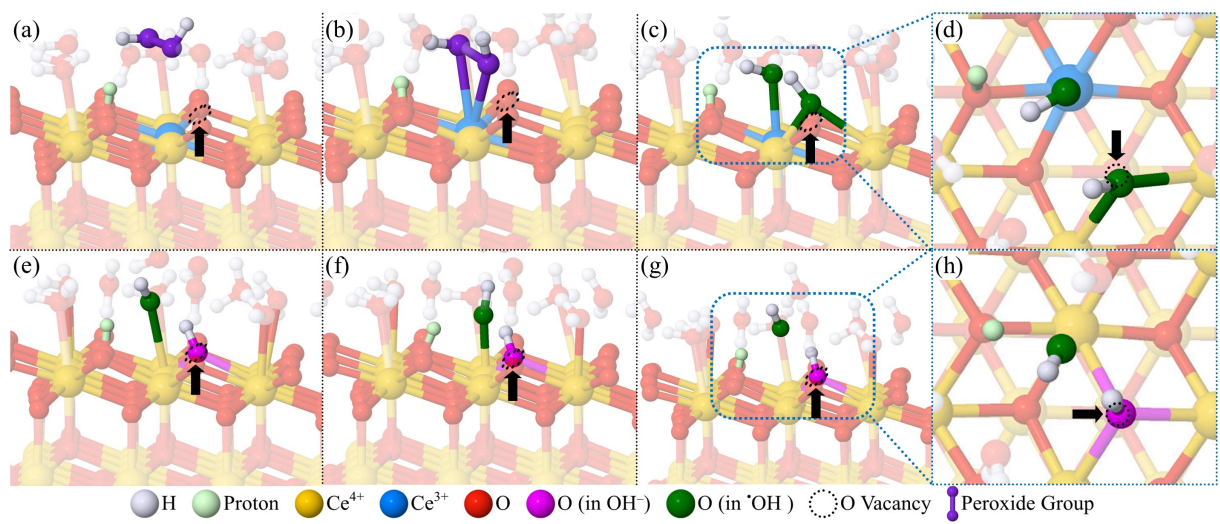

**Figure 5.** Sequential structural changes (a-c, e-g) during spontaneous dissociation of $H_2O_2$ on the defective $CeO_2$ {111} surface (*i.e.*, containing an oxygen vacancy ($V_O^{\bullet\bullet}$)) initially with 1 $Ce^{3+}$ under acidic conditions; (d) and (h) are enlarged top views of enclosed regions in (c) and (g), respectively; black arrows indicate locations of $V_O^{\bullet\bullet}$.

***Summary of CAT biomimetic reaction:*** Based on these results, it is concluded that pH has a significant effect on the interaction between $H_2O_2$ and the defective $CeO_2$ {111} surface. Similar to the SOD biomimetic reaction, the CAT biomimetic reaction rate also is dominated by three factors: (1) the adsorption of $H_2O_2$ on the defective $CeO_2$ {111} surface, (2) the availability of $Ce^{4+}$, and (3) the availability of $OH^-$ ions.

pH plays a significant role in both the adsorption of $H_2O_2$ and the availability of $Ce^{4+}$. When the availabilities of surface $Ce^{4+}$ are the same under both acidic and basic conditions (*i.e.*, defective $CeO_2$ {111} surface with no $Ce^{3+}$), the adsorption of media molecules and $H_2O_2$ under acidic conditions is notably less favorable than under basic conditions since the media species desorb from the surface in the former case (Figures S10 and S11). When the adsorption of $H_2O_2$ on the defective $CeO_2$ {111} surface is similar under both acidic and basic conditions (Geometries 1 and 2, Table 2), the availability of $Ce^{4+}$ is lower under acidic conditions than under basic conditions (*i.e.*, there is one $Ce^{3+}$ for structural stabilization under acidic conditions while there is no $Ce^{3+}$ under basic conditions). Thus, basic pH conditions must provide at least one of the following: (1) more favorable adsorption of $H_2O_2$ or (2) higher availability of $Ce^{4+}$. The most favorable geometries under acidic conditions are Geometries 3 and 4 (Figure 4), in which the $H_2O_2$ spontaneously dissociates by breakage of the O–O bond. However, this is not favorable for promoting the completion of the CAT biomimetic reaction (which occurs by breakage of the O–H bonds in $H_2O_2$) and results in the formation of harmful $^{\bullet}OH$. The implications of this process are discussed subsequently.

(14)

Considering the third factor, the OH⁻ ions required for the CAT biomimetic reaction are derived primarily from two sources: (1) the aqueous environment and (2) the release of protonated surface oxygen (*i.e.*, OH⁻ ions) from the $CeO_2$ surface (which requires an energy of 2.69 eV, Table S1). In terms of the availability of OH⁻ ions, it is clear that basic conditions are more advantageous. Under basic conditions, the abundance of OH⁻ ions in the aqueous environment makes the supply of OH⁻ ions less reliant on the release of protonated surface oxygen from the $CeO_2$ surface. Under acidic conditions, the aqueous environment has a low concentration of OH⁻ ions, so the supply of OH⁻ ions is more reliant on the release of protonated surface oxygen from the $CeO_2$ surface. This results in the formation of $V_O^{\bullet\bullet}$, which cannot be annihilated under acidic conditions owing to the lack of OH⁻ ions; as a result, a newly-formed $V_O^{\bullet\bullet}$ will be present in addition to the initial $V_O^{\bullet\bullet}$. Therefore, under acidic conditions, if the required OH⁻ ions for the CAT biomimetic reaction are supplied by the release of protonated surface oxygen, the concentration of $V_O^{\bullet\bullet}$ ($[V_O^{\bullet\bullet}]$) will increase over time and eventually reach a critical point. The theoretical maximal concentration of intrinsic $V_O^{\bullet\bullet}$ is 25%[47], and further release of protonated surface oxygen to drive the CAT biomimetic reaction would be hindered significantly owing to the energetic unfavorability of the formation of surface defect clusters on $CeO_2$[48] (with a higher formation energy of 4.88 eV compared with 4.46 eV for isolated vacancies, Table S1, Supporting Information).

Overall, it is concluded that the CAT biomimetic reaction is more favorable under basic conditions than under acidic conditions owing to the more favorable adsorption of $H_2O_2$ and the higher availability of $Ce^{4+}$ and OH⁻ ions, which is consistent with previously reported work[5].

**Mechanisms Associated with the pH-Controlled Enzyme-Biomimetic Properties of CeNPs**

Finally, the present results explain the pH-dependence of the enzyme-biomimetic properties of CeNPs. The data demonstrate a significant pH influence on the interaction of both $^{\bullet}O_2^-$ and $H_2O_2$ with the defective $CeO_2$ {111} surface. Considering all of the steps involved in the SOD/CAT cycle, a $V_O^{\bullet\bullet}$ initially is formed by the release of surface oxygen, simultaneously switching two $Ce^{4+}$ to two $Ce^{3+}$ (Step 1, Figures 6 and 7). Under basic conditions, any existing $V_O^{\bullet\bullet}$ are likely to be filled readily by the abundant OH⁻ ions. As a result, the surface structure will be recovered to a state of non-defective-like $CeO_2$ {111} (Step 2, Figure 6). Then, with the introduction of $^{\bullet}O_2^-$ and the subsequent SOD biomimetic reaction, the $Ce^{3+}$ are shifted to $Ce^{4+}$, thus recovering the surface structure to that of non-defective $CeO_2$ {111} (Steps 3 and 4, Figure 6). Subsequently, the CAT biomimetic reaction occurs, with the OH⁻ ions being supplied either from the environment or the release of protonated surface oxygen and reacting with $H_2O_2$ to form two $H_2O$ and $O_2$ (Steps 5 and 6, Figure 6). In this CAT process, the $Ce^{4+}$ are switched to $Ce^{3+}$ owing to the electron transfer from the $H_2O_2$ to the surface. If the OH⁻ ions for the CAT biomimetic reaction are supplied by the release of protonated surface oxygen, the resultant $V_O^{\bullet\bullet}$ will be filled readily once again by the abundant OH⁻ ions, as in Step 2, such that the $CeO_2$ surface is recovered to its original state. This mechanism is consistent with the model proposed by Celardo et al.[49]. Therefore, owing to the reversibility of $Ce^{3+} \leftrightarrow Ce^{4+}$ redox equilibria and the formation ↔ annihilation of $V_O^{\bullet\bullet}$ under basic conditions, the SOD and CAT biomimetic reactions can be performed cyclically, thus scavenging and decomposing ROS to $H_2O$ and $O_2$. This makes $CeO_2$ an excellent antioxidant at basic pH.

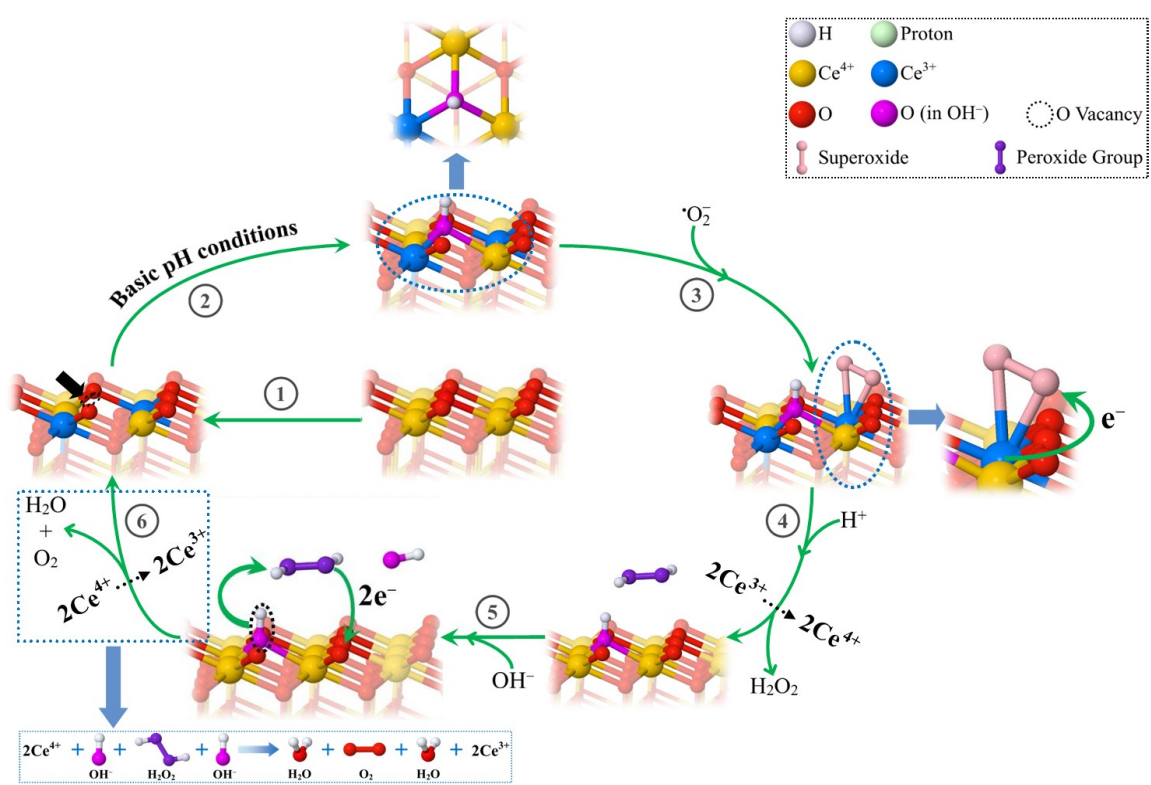

**Figure 6.** Schematic illustration showing changes in surface chemistry of $CeO_2$ driven by SOD and CAT biomimetic reactions under basic conditions; black arrows indicate locations of oxygen vacancies ($V_O^{\bullet\bullet}$); surface state will transfer directly from step (5) to step (2) if both OH⁻ ions are supplied by aqueous environment

Under acidic conditions, the existing $V_O^{\bullet\bullet}$, which is formed in Step 1, can be filled only incompletely by $H_2O$ molecules. This means that, instead of the annihilation of $V_O^{\bullet\bullet}$ as under basic conditions by the abundant OH⁻ ions, the $V_O^{\bullet\bullet}$ still exists under acidic conditions, thereby the $CeO_2$ surface cannot be recovered to a state of non-defective-like $CeO_2$ {111} (Step 2,

(16)

Figure 7). Then, with the introduction of $^\bullet O_2^-$ and the subsequent SOD biomimetic reaction, the $Ce^{3+}$ is shifted to $Ce^{4+}$, but, in contrast to the basic pH cases, one $Ce^{3+}$ is reserved on the surface to stabilize the defective surface structure (Steps 3 and 4, Figure 7). The subsequent CAT biomimetic reaction may occur in one of two different ways (Steps 5, 6, and 7, Figure 7):

(1) If the concentration of $V_O^{\bullet\bullet}$ ($[V_O^{\bullet\bullet}]$) is low, the CAT biomimetic reaction will be likely to proceed similarly to the basic pH cases, with the OH⁻ ions supplied either by the release of protonated surface oxygen or from the aqueous environment. In this former case, new $V_O^{\bullet\bullet}$ are created and the neighboring $Ce^{4+}$ are switched to $Ce^{3+}$ owing to the electron transfer from the $H_2O_2$ to the surface (Step 6, Figure 7).

(2) If the concentration of $V_O^{\bullet\bullet}$ is high, the Fenton biomimetic reaction[50], by breakage of O–O bonds in $H_2O_2$, becomes more likely to occur, with the formation of one $^\bullet OH$ and one OH⁻ ion, which annihilates the $V_O^{\bullet\bullet}$ (Step 7, Figure 7):

$$H_2O_2 + Ce^{3+} \rightarrow Ce^{4+} + {}^\bullet OH + OH^- \qquad (4)$$

Step 6 successfully completes the CAT biomimetic reaction and decomposes $H_2O_2$ to $H_2O$ and $O_2$. However, as discussed previously, the source of OH⁻ ions under acidic conditions is likely to be dominated by the release of protonated surface oxygen, which increases the concentration of $V_O^{\bullet\bullet}$ and $Ce^{3+}$. Thus, Step 6 will be limited significantly when the surface concentration of $V_O^{\bullet\bullet}$ reaches a critical point owing to the energetic unfavorability of the formation of surface defect clusters on $CeO_2$. A high concentration of $V_O^{\bullet\bullet}$ reduces the stability of the surface structure; therefore, in order to stabilize the surface structure with a high concentration of $V_O^{\bullet\bullet}$, Step 7 will occur simultaneously with Step 6 to annihilate the $V_O^{\bullet\bullet}$ by the accommodation of the OH⁻ ions formed in the Fenton biomimetic reaction. Thus, through Step 7, the $CeO_2$ surface can be recovered to a state of non-defective $CeO_2$ {111}, which confers on $CeO_2$ the capacity to perform the SOD and CAT biomimetic reactions cyclically. Thus, under acidic conditions, the Fenton biomimetic reaction occurs simultaneously with the SOD and CAT biomimetic reactions in this cycle, forming destructive $^\bullet OH$. This makes $CeO_2$ a harmful prooxidant at acidic pH.

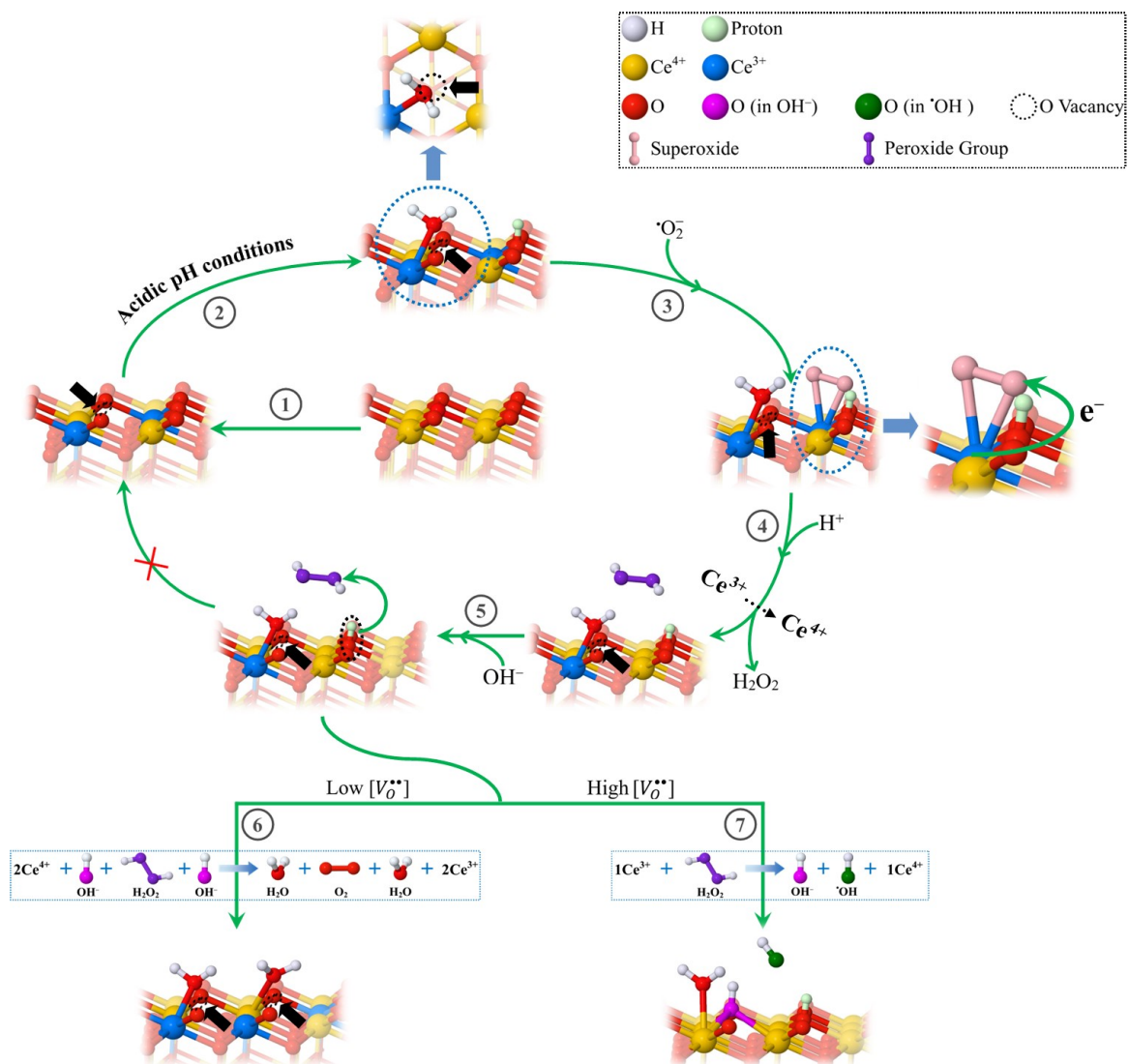

**Figure 7.** Schematic illustration showing changes in surface chemistry of $CeO_2$ driven by SOD and CAT biomimetic reactions under acidic conditions; black arrows indicate locations of oxygen vacancies ($V_O^{\bullet\bullet}$)

*Summary of mechanism:* The pH has an insignificant influence on the SOD biomimetic reaction induced by CeNPs; however, in contrast, the pH affects the CAT biomimetic reaction in distinct ways. Under basic conditions, owing to the abundant $OH^-$ ions from the environment and the preservation of the recovery of the $CeO_2$ surface structure, the CAT biomimetic reaction primarily proceeds by decomposition of $H_2O_2$ to harmless $H_2O$ and $O_2$. Under acidic conditions, owing to the lack of $OH^-$ ions in the environment and the imperfect reversibility of $Ce^{3+} \leftrightarrow Ce^{4+}$ redox equilibria and the formation $\leftrightarrow$ annihilation of $V_O^{\bullet\bullet}$, the CAT biomimetic reaction, decomposing $H_2O_2$ to harmless $H_2O$ and $O_2$, is limited, which may result in an accumulation of $H_2O_2$. Furthermore, a Fenton biomimetic reaction is likely to occur simultaneously, resulting in the formation of $^{\bullet}OH$. The presence of excess $H_2O_2$ also can lead to the generation of $^{\bullet}OH$ through a metal-catalyzed (with metal elements in blood) Fenton

(18)

reaction[51]. Therefore, the production of harmful ROS ($^\bullet$OH and $H_2O_2$) under acidic conditions is significantly higher than that under basic conditions. This is consistent with reported experimental work which demonstrates that CeNPs can facilitate the decomposition of $H_2O_2$ to $^\bullet$OH and that an increase of pH from 3.0 to 9.0 decreases the yield of the $^\bullet$OH[52, 53]. It is known widely that $^\bullet$OH is the most destructive ROS *in vivo*, causing severe oxidative damage to cells[51]. Therefore, under acidic conditions, $CeO_2$ exhibits cytotoxicity due to the accumulation of $H_2O_2$ and the generation of $^\bullet$OH.

**Conclusions**

DFT calculations have been used to interpret the pH-induced behavior of the stable $CeO_2$ {111} surface. Results indicate that pH plays a critical role in the biomimetic reactions induced by CeNPs. Under basic conditions, the SOD and CAT biomimetic reactions can be performed cyclically, causing CeNPs to act as antioxidants, eliminating and decomposing harmful ROS to $H_2O$ and $O_2$. In contrast, under acidic conditions, $H_2O_2$ may accumulate and a Fenton biomimetic reaction occurs simultaneously with the SOD and CAT biomimetic reactions, thereby resulting in the generation of destructive $^\bullet$OH and causing CeNPs to act as cytotoxic prooxidants. The combination of pH-dependent antioxidative and prooxidative properties confers on CeNPs tremendous potential as an effective treatment for cancer.

**Supporting Information**

Calculated formation energies of an oxygen vacancy ($V_O^{\bullet\bullet}$) on pristine $CeO_2$ {111} surface by different mechanisms; total energy as a function of size of *k*-point mesh for $CeO_2$ {111} slab supercell; initial and optimized geometries with different arrangements of media molecules and ROS adsorbed on the defective $CeO_2$ {111} surface; different configurations of the defective $CeO_2$ {111} surface; sequential structural changes during optimization when two $OH^-$ ions and an $H_2O_2$ are introduced above the defective $CeO_2$ {111} surface.

**Corresponding Authors**

**Hongyang Ma** − School of Materials Science and Engineering, UNSW Sydney, Sydney, NSW 2052, Australia; hongyang.ma@unsw.edu.au

**Judy N. Hart** − School of Materials Science and Engineering, UNSW Sydney, Sydney, NSW 2052, Australia; j.hart@unsw.edu.au

**Authors**

**Zhao Liu** − Sino-French Institute of Nuclear Engineering and Technology, Sun Yat-sen University, Zhuhai 519082, China

**Pramod Koshy** − School of Materials Science and Engineering, UNSW Sydney, Sydney, NSW 2052, Australia

**Charles C. Sorrell** − School of Materials Science and Engineering, UNSW Sydney, Sydney, NSW 2052, Australia

**Notes**

The authors declare that they have no conflict of interest.

**Acknowledgments**

This work was financially supported by the Australian Research Council Discovery Project scheme. This research was undertaken with the assistance of computational resources provided by the Australian Government through the National Computational Infrastructure (NCI) under the National Computational Merit Allocation Scheme.

**TOC Graphic**

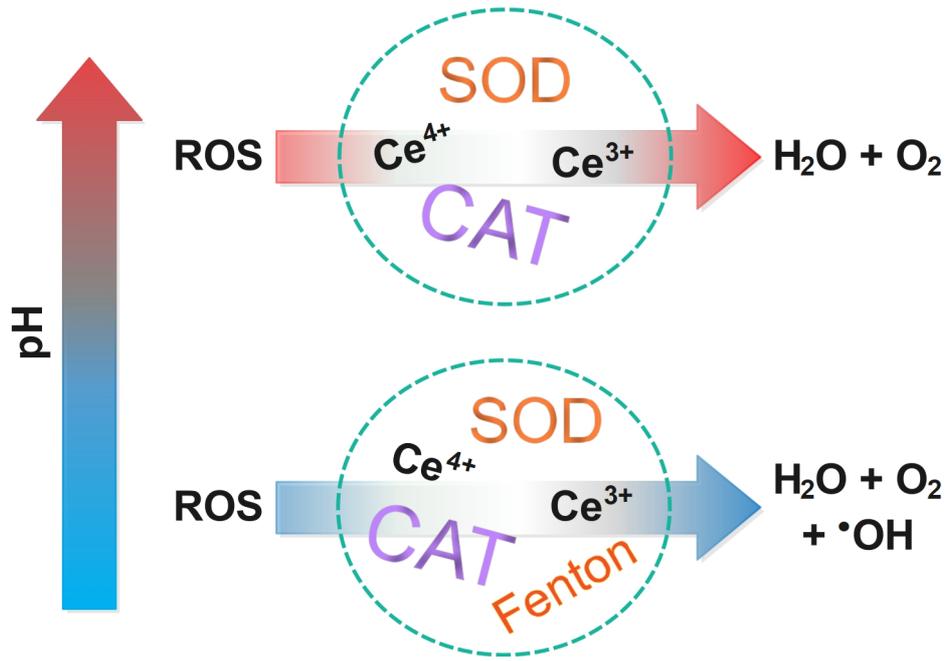

(25)